\documentclass[twoside]{LCWS11}
\usepackage[latin1]{inputenc}
\usepackage[dvips]{graphicx,epsfig,color}
\usepackage{wrapfig,rotating}
\usepackage{amssymb,amsmath,array}
\usepackage{cite}
\begin{document}
\title{
Pair-production and Three-body Decay of the Lighter Stop at the ILC
in One-loop Order in the MSSM} 
\author{M. Jimbo$^1$, T. Inoue$^2$, T. Jujo$^2$, T. Kon$^2$, 
T. Ishikawa$^3$, Y. Kurihara$^3$, K. Kato$^4$ and M. Kuroda$^5$
\vspace{.3cm}\\
1- Chiba University of Commerce \\
Ichikawa, Chiba - Japan
\vspace{.1cm}\\
2- Seikei University \\
Musashino, Tokyo - Japan
\vspace{.1cm}\\
3- KEK (Tsukuba) \\
Tsukuba, Ibaraki - Japan
\vspace{.1cm}\\
4- Kogakuin University \\
Shinjuku, Tokyo - Japan
\vspace{.1cm}\\
5- Meiji Gakuin University \\
Yokohama, Kanagawa - Japan\\
}

\maketitle

\begin{abstract}
We have been developing a program package called {\tt GRACE/SUSY-loop} which automatically
calculates the MSSM amplitudes in one-loop order. We present numerical results of calculations for
pair-production and three-body decay of the lighter stop ($\widetilde{t}_1$) at the International Linear Collider (ILC)
using {\tt GRACE/SUSY-loop}.
Since the distributions of missing transverse energy (MET) depend on mass spectrum of SUSY particles,
we consider two scenarios on three-body decay of $\widetilde{t}_1$.  In these scenarios,
both QCD and EW corrections have positive sign for decay widths and cross sections.
\end{abstract}

\section{Introduction}
Supersymmetry (SUSY) between bosons and fermions at the unification-energy scale is considered
one of the most promising extension of the standard model (SM) of particle physics.  Among
the supersymmetric theories, the minimal supersymmetric extension of the SM (MSSM) is a
well studied framework of SUSY.  Since SUSY is a broken symmetry at the electroweak-energy scale,
its relic is expected to remain as a rich spectrum of heavy SUSY particles, which makes calculation
of amplitudes very complicated.

For many-body final states, each production process or decay process is described by
a large number of Feynman diagrams even in tree-level order.  There are still more
Feynman diagrams in one-loop order even for two-body final states.  For this reason,
we have developed the {\tt GRACE} system~\cite{Yuasa:1999rg}, which enables us to calculate
amplitudes automatically.  A version of the {\tt GRACE} system called {\tt GRACE/SUSY-loop}~\cite
{Fujimoto:2007bn} is a program package for automated computations of the MSSM in one-loop order.
Other groups independently developed program packages {\tt SloopS}~\cite{Baro:arXiv0906.1665} and
{\tt FeynArt/Calc}~\cite{Hahn:2000jm} for automatic calculations of the MSSM amplitudes in one-loop order.

Recently, we have calculated radiative corrections of production processes and decay
processes of SUSY particles in the framework of the MSSM using
{\tt GRACE/SUSY-loop}~\cite{Fujimoto:2007bn, Iizuka:2010bh, Jimbo:2010}.
In this paper, we show numerical results of the MET distributions for pair-production
and three-body decay of $\widetilde{t}_1$ at the ILC in tree-level order,
and also show those of the radiative corrections to cross sections for pair-production~\cite{Eberl:2010}
and decay widths for three-body decay of $\widetilde{t}_1$~\cite{Guasch:2002ez, Eberl:2010decay} in one-loop order.

\section{MET distributions}
In the experiments at the Large Hadron Collider (LHC), searches for SUSY particles are performed using an
experimental cut on MET as is required to be larger than 100 GeV~\cite{ATLAS:2011}
because of background-events avoidance.  When a SUSY particle is heavy and decays directly to a charged particle
and the lightest SUSY particle (LSP), e.g. $\widetilde{t}_1 \rightarrow t + \widetilde{\chi}_1^0$, the MET
is expected to be large because the LSP is not observed in detectors and the momentum of the parent particle is small.
It, however, can be smaller if a SUSY particle is light and has a large momentum because momenta of particles produced
from the decay are boosted along that of the parent particle.  Although no light SUSY particle has been
observed at the LHC, there is room for it to survive if MET is small.

\subsection{Two scenarios}
Since the Yukawa coupling of top quark is large, the mass of $\widetilde{t}_1$ can be smaller than
that of other squarks.  So here we focus on $\widetilde{t}_1$ pair-production $e^- + e^+
\rightarrow \widetilde{t}_1 + \widetilde{t}_1^*$ and decay at the ILC.  To investigate
the possibility that $M_{\widetilde{t}_1}$ is small enough, we set the mass range of $\widetilde{t}_1$
below the threshold of the two-body decay.  Since the MET distributions depend on mass spectrum of
SUSY particles, we consider two scenarios for three-body decay of $\widetilde{t}_1$ as follows:
\begin{description}
\item{\bf Scenario 1.} Large slepton masses\\
The major decay mode is $\widetilde{t}_1 \rightarrow b W^+ \widetilde{\chi}_1^0$~\cite{Iizuka:2010bh, Jimbo:2010},
and its branching ratio is nearly 100\% if the mass range of $\widetilde{t}_1$ is between the threshold
of the three-body decay, $M_{\widetilde{t}_1} = 279$ GeV and the lowest threshold of the two-body decay, which is for the
mode $\widetilde{t}_1 \rightarrow t + \widetilde{\chi}_1^0$, $M_{\widetilde{t}_1} = 368$ GeV.
\item{\bf Scenario 2.} Small slepton masses\\
`Semi-Leptonic' decay modes, $\widetilde{t}_1 \rightarrow b l^+ \widetilde{\nu}_l$ and $\widetilde{t}_1
\rightarrow b \widetilde{l}^+ \nu_l ~(l=e,\mu,\tau)$, dominate, and the branching ratio of the mode 
$\widetilde{t}_1 \rightarrow b W^+ \widetilde{\chi}_1^0$ is negligible small due to the small phase space.
The lowest threshold of the two-body decay, which is for the mode $\widetilde{t}_1 \rightarrow b + \widetilde{\chi}_1^+$,
is $M_{\widetilde{t}_1} = 298$ GeV.
\end{description}

We set SUSY parameters for these scenarios as in Table~\ref{tab:param}.
\begin{table}[htb]
\centerline{\small \begin{tabular}{|cc|cc|cc|cc|}
\hline
\multicolumn{4}{|l|}{\bf Scenario 1} & \multicolumn{4}{l|}{\bf Scenario 2}\\ \hline
$\tan\beta$ & 10 & $M_{\widetilde{t}_1}$ & {\it variable} & $\tan\beta$ & 7 & $M_{\widetilde{t}_1}$ & {\it variable}\\
$\mu$ & -750 GeV & $M_{\widetilde{t}_2}$ & 480 GeV & $\mu$ & -500 GeV & $M_{\widetilde{t}_2}$ & 600 GeV\\
$M_2$ & 400 GeV & $\theta_t$ & $0.8\pi$ & $M_2$ & 300 GeV & $\theta_t$ & $0.8\pi$\\
$M_{\widetilde{l}_1^+}$ & 325 GeV & $M_{\widetilde{b}_1}$ & 330 GeV & $M_{\widetilde{l}_1^+}$ & 170 GeV & $M_{\widetilde{b}_1}$ & 330 GeV\\
$M_{\widetilde{l}_2^+}$ & 370 GeV & $\theta_b$ & $0.6\pi$ & $M_{\widetilde{l}_2^+}$ & 175 GeV & $\theta_b$ & $0.6\pi$\\
$\theta_{e,\mu}$ & $0.05\pi$ & $M_A$ & 525 GeV & $\theta_{e,\mu}$ & $0.01\pi$ & $M_A$ & 300 GeV\\
$\theta_\tau$ & $0.2\pi$ & $M_{\widetilde{g}}$ & 1389 GeV & $\theta_\tau$ & $0.2\pi$ & $M_{\widetilde{g}}$ & 1042 GeV\\
$M_{\widetilde{\nu}_{e,\mu}}$ & 316 GeV & $M_{\widetilde{\chi}_1^0}$ & 194 GeV & $M_{\widetilde{\nu}_{e,\mu}}$ & 151 GeV & $M_{\widetilde{\chi}_1^0}$ & 146 GeV\\
$M_{\widetilde{\nu}_\tau}$ & 328 GeV & $M_{\widetilde{\chi}_1^+}$ & 396 GeV & $M_{\widetilde{\nu}_\tau}$ & 152 GeV & $M_{\widetilde{\chi}_1^+}$ & 294 GeV\\
\hline
\end{tabular}}
\caption{SUSY parameters for two scenarios.}
\label{tab:param}
\end{table}

\subsection{Numerical results of MET in tree-level order}
\begin{figure}[htb]
\begin{tabular}{ccc}
\begin{minipage}{0.2\columnwidth}
\begin{center}
\end{center}
\end{minipage}
\begin{minipage}{0.5\columnwidth}
\begin{center}
\includegraphics[width=\columnwidth]{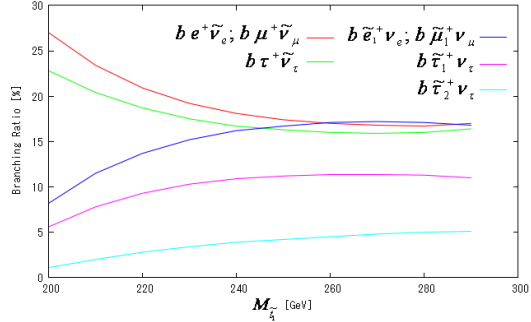}
\end{center}
\end{minipage}
\begin{minipage}{0.2\columnwidth}
\begin{center}
\end{center}
\end{minipage}
\end{tabular}
\caption{$M_{\widetilde{t}_1}$ dependence of branching ratios for three-body decay of $\widetilde{t}_1$ in tree-level order
according to Scenario 2.}
\label{Fig:BRtree}
\end{figure}

In Scenario 1, $BR(\widetilde{t}_1 \rightarrow b W^+ \widetilde{\chi}_1^0) \simeq 1$ in the range of $M_{\widetilde{t}_1}$
for which three-body decay of $\widetilde{t}_1$ is dominant as studied in \cite{Iizuka:2010bh, Jimbo:2010}.  In Scenario 2,
however, there are several `Semi-Leptonic' decay modes.  Figure~\ref{Fig:BRtree} shows numerical results of
$M_{\widetilde{t}_1}$ dependence of branching ratios for three-body decay of $\widetilde{t}_1$ in tree-level order
according to Scenario 2.

\begin{figure}[hbt]
\begin{minipage}{0.49\columnwidth}
\begin{center}
\includegraphics[width=\columnwidth]{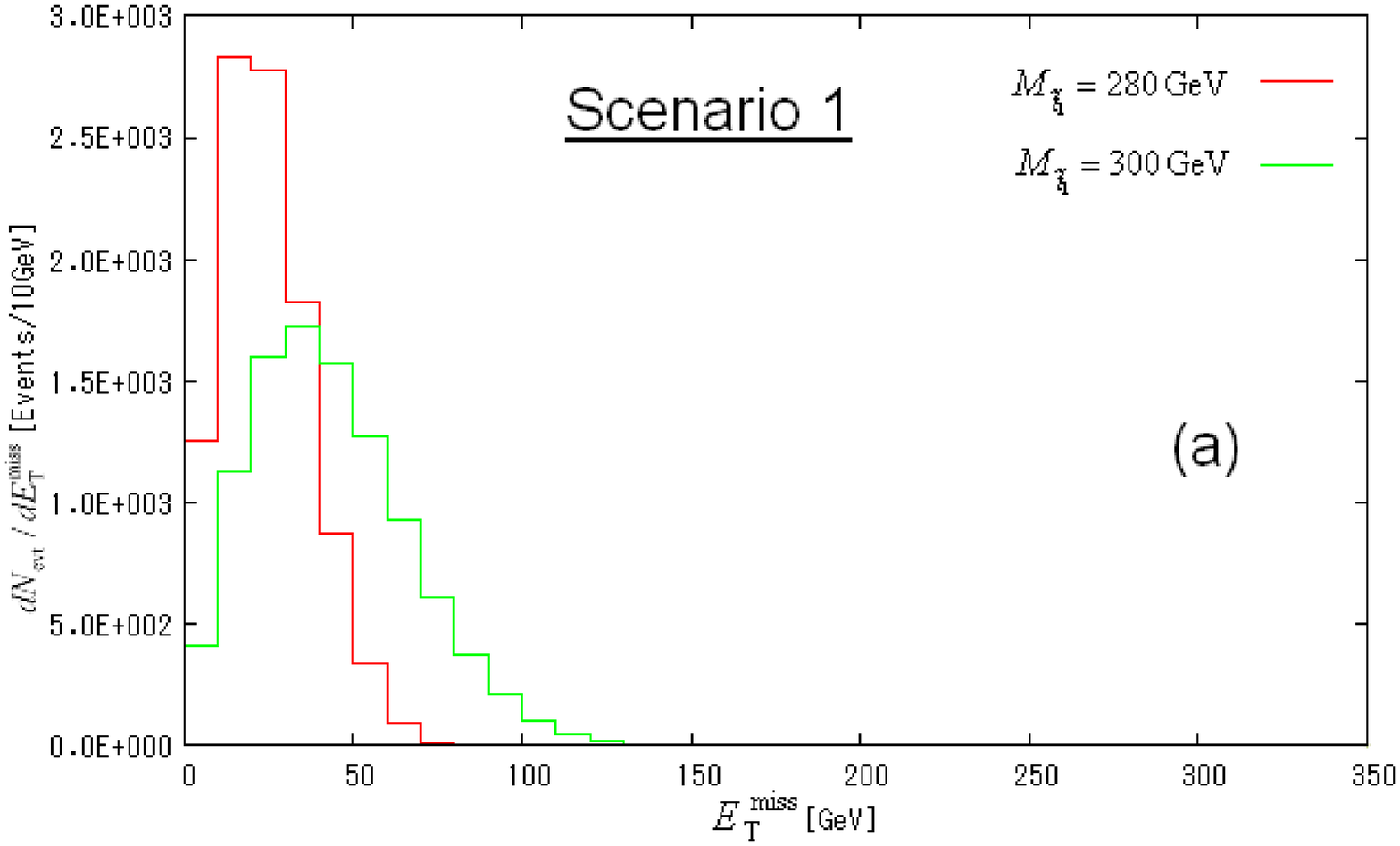}
\end{center}
\end{minipage}
\begin{minipage}{0.49\columnwidth}
\begin{center}
\includegraphics[width=\columnwidth]{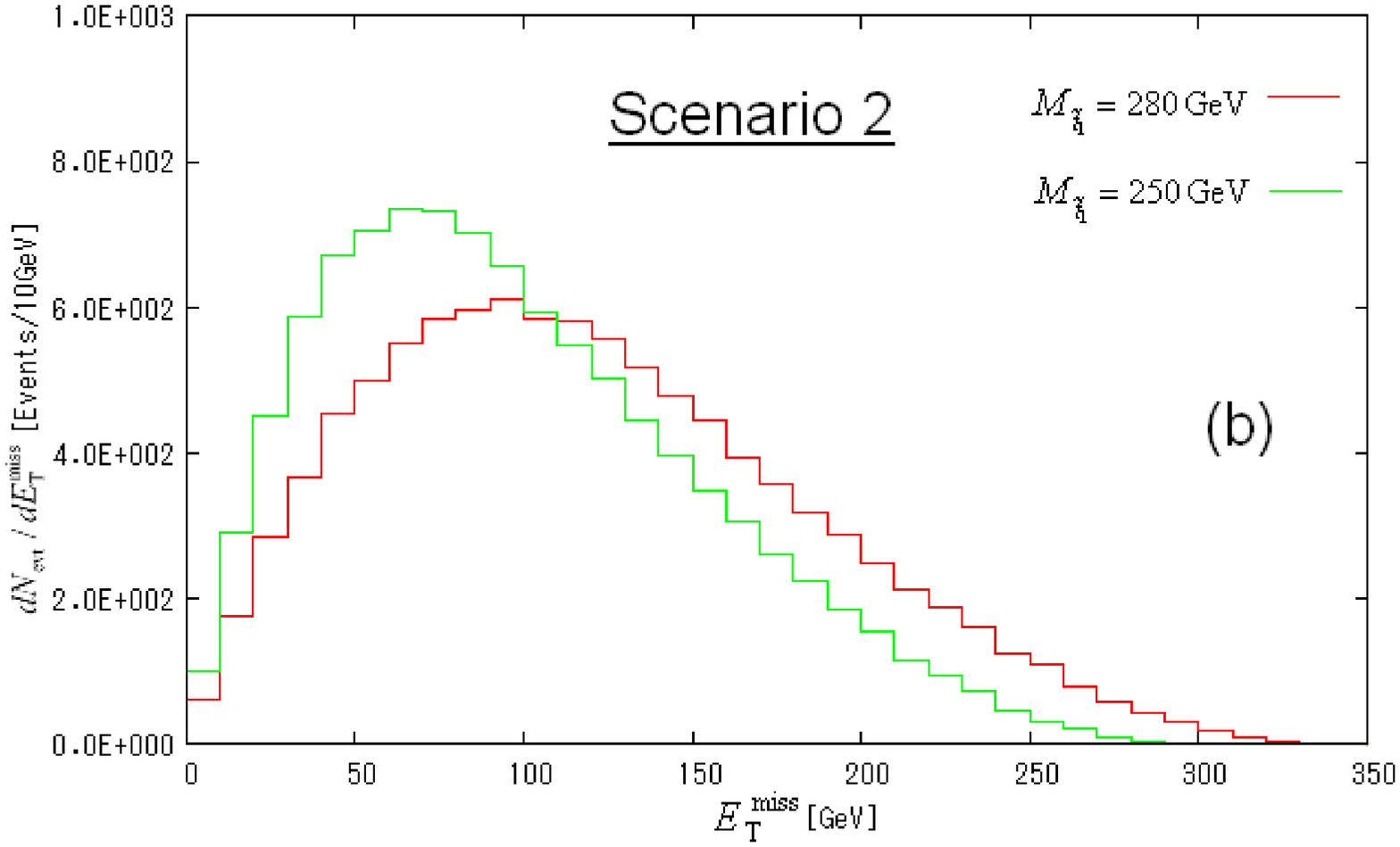}
\end{center}
\end{minipage}
\caption{MET distributions at $\sqrt{s}=1$ TeV in tree-level order. (a) for $e^- + e^+ \rightarrow \widetilde{t}_1 +
\widetilde{t}_1^* \rightarrow (b W^+ {\widetilde{\chi}_1^0}) + (\bar{b} W^- {\widetilde{\chi}_1^0})$ according to Scenario 1;
and (b) for $e^- + e^+ \rightarrow \widetilde{t}_1 + \widetilde{t}_1^* \rightarrow (b e^+ \widetilde{\nu}_e) +
(\bar{b} e^- \widetilde{\nu}_e^*)$ according to Scenario 2.  Red lines indicate $M_{\widetilde{t}_1} = 280$ GeV.
Green lines in (a) and (b) indicate $M_{\widetilde{t}_1} = 300$ GeV and 250 GeV, respectively.}
\label{Fig:MET}
\end{figure}

Figure~\ref{Fig:MET} shows numerical results of (a) MET distributions for $e^- + e^+ \rightarrow \widetilde{t}_1 +
\widetilde{t}_1^* \rightarrow (b W^+ {\widetilde{\chi}_1^0}) + (\bar{b} W^- {\widetilde{\chi}_1^0})$ according to
Scenario 1 and (b) those for $e^- + e^+ \rightarrow \widetilde{t}_1 + \widetilde{t}_1^* \rightarrow
(b e^+ \widetilde{\nu}_e) + (\bar{b} e^- \widetilde{\nu}_e)$ according to Scenario 2, respectively, at $\sqrt{s}=1$ TeV
in tree-level order.  In our calculations, events are generated by {\tt SPRING}~\cite{Kawabata:1986} which is a
built-in package in the {\tt GRACE} system.  In Scenario 1, the MET is calculated for the two LSP's.  In Scenario 2,
a major decay mode $\widetilde{t}_1 \rightarrow b \, e^+ \, \widetilde{\nu}_e$ is selected, and the MET is calculated for the
two sneutrinos.

\begin{figure}[htbp]
\centerline{\includegraphics[width=0.85\columnwidth]{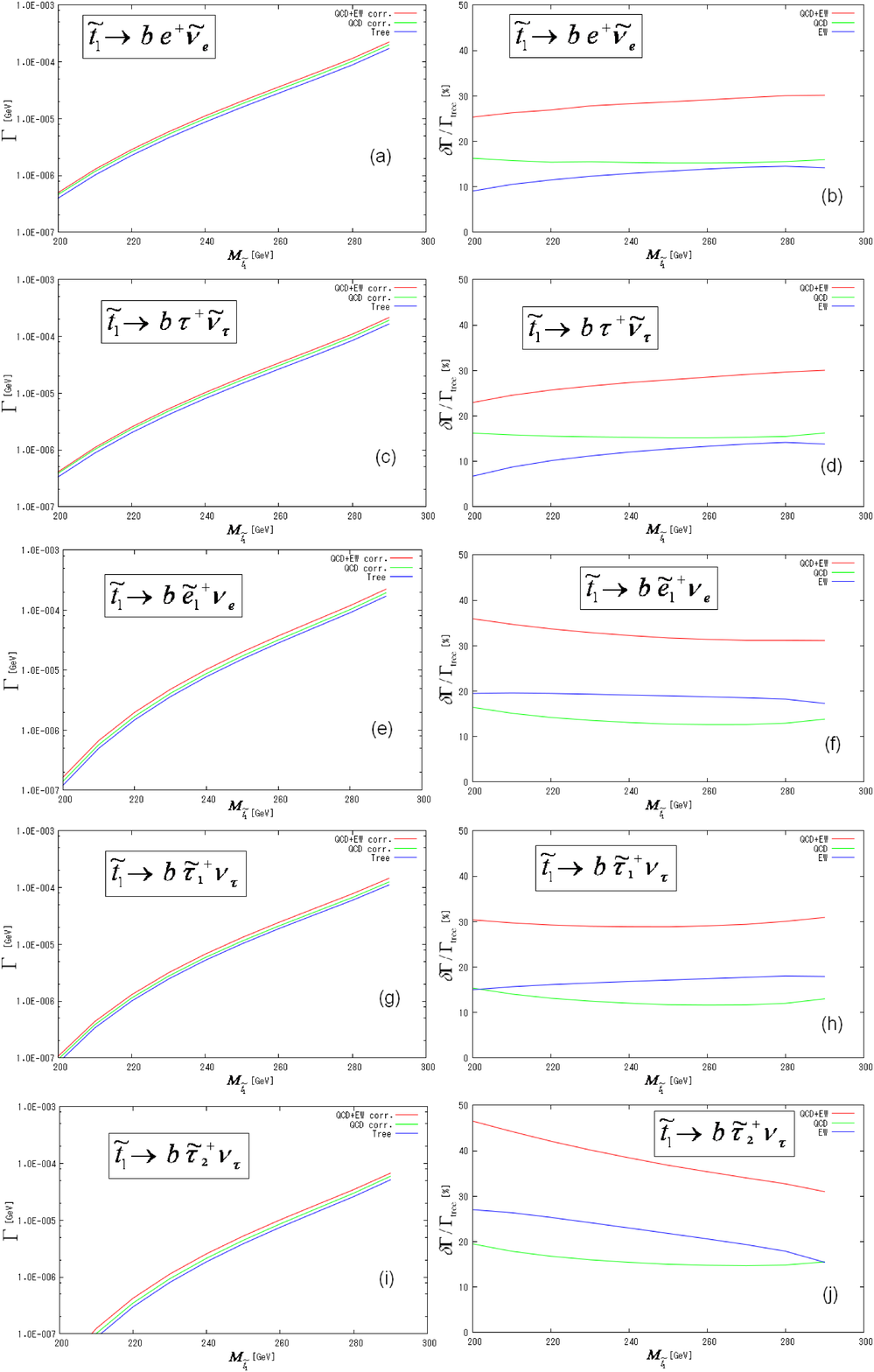}}
\caption{$M_{\widetilde{t}_1}$ dependence of decay widths for three-body decay processes of $\widetilde{t}_1$ and their
correction rates according to Scenario 2.}
\label{Fig:decay}
\end{figure}

We should note that the MET in Scenario 1 is small because the difference of numerical values between $M_{\widetilde{t}_1}$
and the sum of $M_W$ and $M_{\widetilde{\chi}_1^0}$ is small.  Although each W boson decays to a pair of quarks or that of
a charged lepton and a neutrino, resultant MET distributions are similar to Figure~\ref{Fig:MET} (a) except for the
normalization factors because the momenta of the final particles from the W-boson decay are boosted along that of
$\widetilde{t}_1$.  It is also notable that the peaks of the MET distributions are less than 100 GeV in Scenario 2.

\section{Numerical results of radiative corrections in one-loop order}
In Scenario 1, the radiative corrections to decay widths of $\widetilde{t}_1$ in one-loop order have already been studied
in \cite{Iizuka:2010bh, Jimbo:2010}.  In Scenario 2, decay widths for several `Semi-Leptonic' decay modes of
$\widetilde{t}_1$ are in the same order.  Figure~\ref{Fig:decay} shows numerical results of $M_{\widetilde{t}_1}$ dependence
of decay widths of $\widetilde{t}_1$ and their correction rates according to Scenario 2.  For the decay modes
$\widetilde{t}_1 \rightarrow b l^+ \widetilde{\nu}_l$ [(a) - (d) in Figure~\ref{Fig:decay}], $\delta\Gamma_{\rm QCD} >
\delta\Gamma_{\rm EW} > 0$, and for the decay modes $\widetilde{t}_1 \rightarrow b \widetilde{l}^+ \nu_l$ [(e) - (j) in
Figure~\ref{Fig:decay}], $\delta\Gamma_{\rm EW} > \delta\Gamma_{\rm QCD} > 0$.  The difference in the behavior of
$\delta\Gamma_{\rm EW}$ between the two is due to that of the QED corrections.  Each decay width is enhanced by the total
radiative correction in one-loop order.

\begin{figure}[hbt]
\begin{tabular}{ccc}
\begin{minipage}{0.2\columnwidth}
\begin{center}
\end{center}
\end{minipage}
\begin{minipage}{0.5\columnwidth}
\begin{center}
\includegraphics[width=\columnwidth]{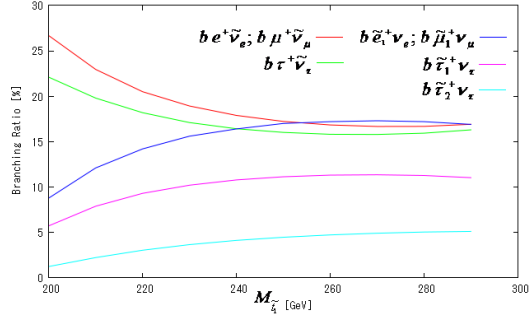}
\end{center}
\end{minipage}
\begin{minipage}{0.2\columnwidth}
\begin{center}
\end{center}
\end{minipage}
\end{tabular}
\caption{$M_{\widetilde{t}_1}$ dependence of branching ratios for three-body decay of $\widetilde{t}_1$ in one-loop order
according to Scenario 2.}
\label{Fig:BRloop}
\end{figure}

Figure~\ref{Fig:BRloop} shows numerical results of $M_{\widetilde{t}_1}$ dependence of branching ratios for three-body
decay of $\widetilde{t}_1$ in one-loop order according to Scenario 2.  The results are similar to those in tree-level order
(Figure~\ref{Fig:BRtree}) because $\delta\Gamma$'s of the decay modes are in the same order.

\begin{figure}[htb]
\begin{minipage}{0.49\columnwidth}
\begin{center}
\includegraphics[width=\columnwidth]{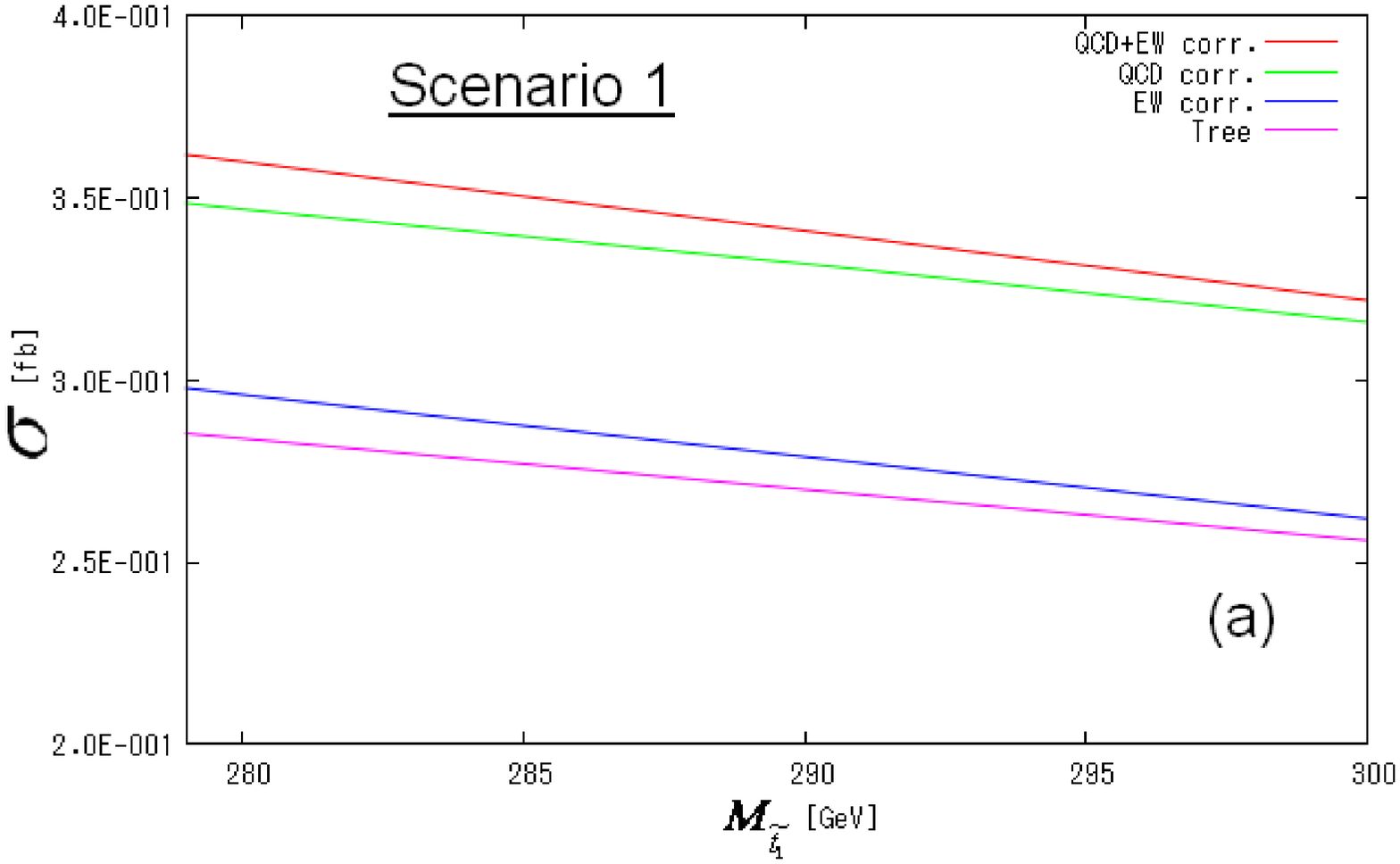}
\end{center}
\end{minipage}
\begin{minipage}{0.49\columnwidth}
\begin{center}
\includegraphics[width=\columnwidth]{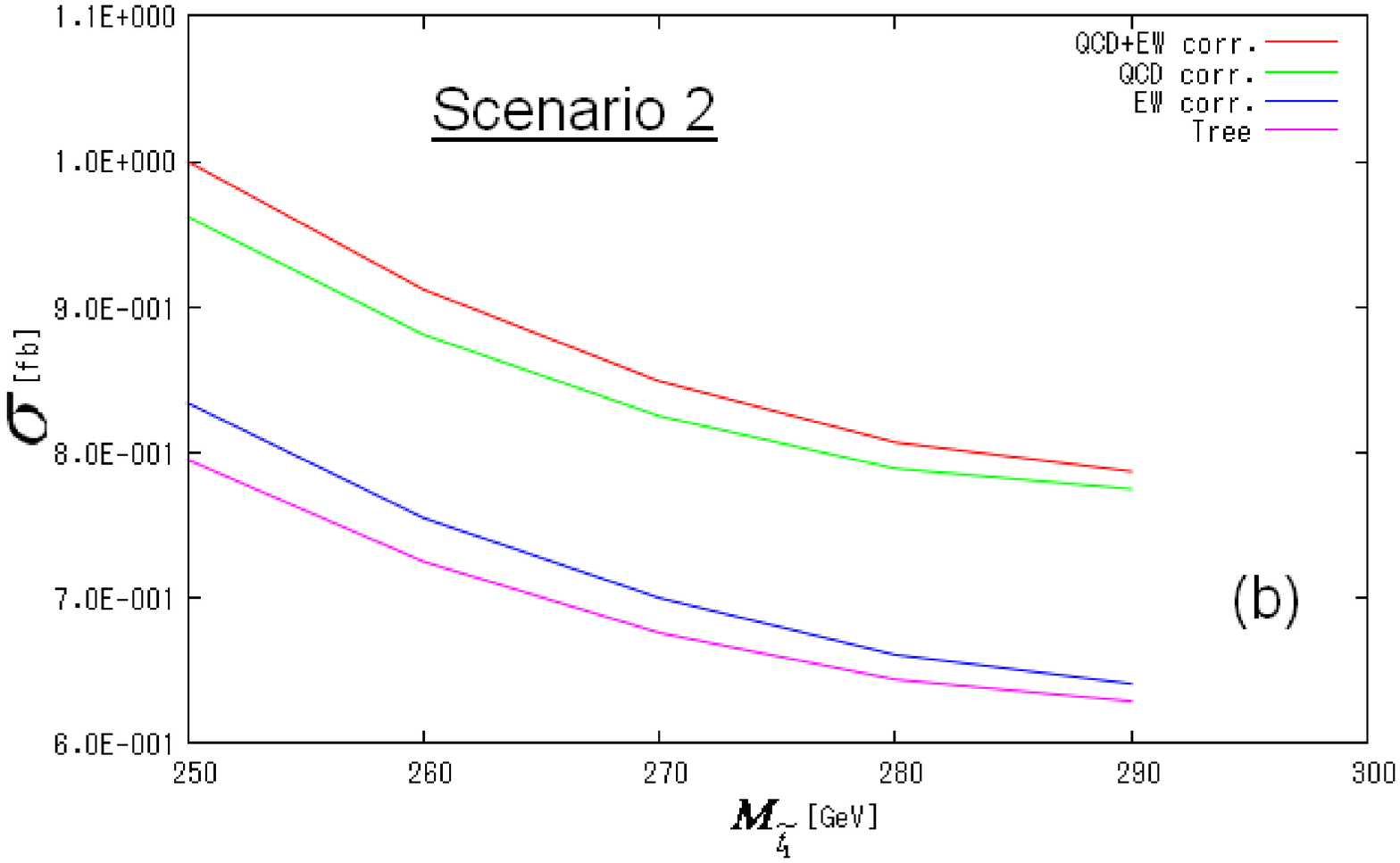}
\end{center}
\end{minipage}
\caption{$M_{\widetilde{t}_1}$ dependence of cross sections at $\sqrt{s}=1$ TeV. (a) for $e^- + e^+ \rightarrow
\widetilde{t}_1 + \widetilde{t}_1^* \rightarrow (b W^+ {\widetilde{\chi}_1^0}) + (\bar{b} W^- {\widetilde{\chi}_1^0})
\rightarrow (b e^+ \nu_e {\widetilde{\chi}_1^0}) + (\bar{b} e^- \bar{\nu}_e {\widetilde{\chi}_1^0})$ according to Scenario 1; and (b) for $e^- + e^+ \rightarrow\widetilde{t}_1 + \widetilde{t}_1^* \rightarrow (b e^+ \widetilde{\nu}_e) + (\bar{b} e^-
\widetilde{\nu}_e^*)$ according to Scenario 2.}
\label{Fig:CS}
\end{figure}

Figure~\ref{Fig:CS} shows numerical results of (a) $M_{\widetilde{t}_1}$ dependence of cross sections for the
$\widetilde{t}_1$-pair production and succeeding decay process, $e^- + e^+ \rightarrow \widetilde{t}_1 + \widetilde{t}_1^* \rightarrow
(b W^+ {\widetilde{\chi}_1^0}) + (\bar{b} W^- {\widetilde{\chi}_1^0}) \rightarrow (b e^+ \nu_e {\widetilde{\chi}_1^0})
+ (\bar{b} e^- \bar{\nu}_e {\widetilde{\chi}_1^0})$ according to Scenario 1 and (b) that for the $\widetilde{t}_1$-pair
production and succeeding decay process, $e^- + e^+ \rightarrow \widetilde{t}_1 + \widetilde{t}_1^* \rightarrow (b e^+ \widetilde{\nu}_e)
+ (\bar{b} e^- \widetilde{\nu}_e)$, according to Scenario 2 at $\sqrt{s}=1$ TeV.  For comparison, a decay mode $W^+ \rightarrow
e^+ \, \widetilde{\nu}_e$ is selected in Scenario 1, and a decay mode $\widetilde{t}_1 \rightarrow b \, e^+ \, \widetilde{\nu}_e$
is selected in Scenario 2.   In both scenarios, $\delta\sigma_{\rm QCD} > \delta\sigma_{\rm EW} > 0$, and cross sections are enhanced
by the total radiative corrections in one-loop order.

We should note that the MET distributions in one-loop order are similar to those in tree-level order except for
the normalization factors, and that the MET distributions in the two scenarios can be distinguished by the
$b \, \bar{b} \, e^- \, e^+$ tagging.

\section{Summary}
We have calculated the MET distributions for the pair-production and three-body decay of $\widetilde{t}_1$ in tree-level order,
and also calculated the radiative corrections in one-loop order of the MSSM for two scenarios on slepton masses.
We found that the peak of the MET distributions can be less than 100 GeV if $M_{\widetilde{t}_1} \le 300$ GeV in both scenarios,
so events for ${\rm MET} < 100$ GeV should be also analyzed in detail for ILC study.  In the two scenarios, both QCD and EW
corrections in one-loop order have positive sign for decay widths and cross sections in the range of $M_{\widetilde{t}_1}$ for which
three-body decay of $\widetilde{t}_1$ is dominant.

\section*{Acknowledgments}
This work was supported in part by Grant-in-Aid for Scientific Research (B)(20340063) and
(C)(23540328).


\begin{footnotesize}


\end{footnotesize}


\end{document}